\documentclass[aps,prl,twocolumn,showkeys,floatfix]{revtex4}
\usepackage{graphicx}
\usepackage{amsmath}%
\usepackage{amsfonts}%
\usepackage{amssymb}

\begin{document}
	  
\title{Violation of the Zeroth Law of Thermodynamics in Systems with 
\\ Negative Specific Heat}

\author{A. Ram{\'i}rez-Hern{\'a}ndez$^{1,2}$}
\author{H. Larralde$^{2}$}
\author{F. Leyvraz$^{2}$}

\affiliation{$^{1}$Facultad de Ciencias, Universidad Aut{\'o}noma del
Estado de Morelos, C.P. 62210. Cuernavaca, Morelos, M\'exico.}

\affiliation{$^{2}$Instituto de Ciencias F{\'i}sicas, Universidad
Nacional Aut{\'o}noma de M{\'e}xico, C.P. 62251. Apartado Postal 48-3,
Cuernavaca, Morelos, M\'exico.}


\begin{abstract}
We show that systems with negative specific heat can violate the
zeroth law of thermodynamics. By both numerical simulations and by
using exact expressions for free energy and microcanonical entropy it
is shown that if two systems with the same intensive parameters but
with negative specific heat are thermally coupled, they undergo a
process in which the total entropy increases irreversibly. The final
equilibrium is such that two phases appear, that is, the subsystems have
different magnetizations and internal energies at temperatures which are
equal in both systems, but that can be different from the initial
temperature.
\end{abstract}
 
\maketitle
 
The zeroth law of thermodynamics is among the most fundamental
assumptions concerning macroscopic systems in equilibrium. In its
textbook form, it concerns three systems $A$, $B$ and $C$ at
equilibrium, that is, such that their macroscopic variables are
time-independent: if $A$ and $B$ are in thermal equilibrium with each
other, that is, if no heat flow arises when they are brought into
thermal contact, and if further system $B$ is similarly in thermal
equilibrium with system $C$, then the zeroth law states that system
$A$ will also be in equilibrium with system $C$. By bringing two
systems in thermal contact, one means allowing a weak coupling through
which energy can be exchanged between the systems.  From the above
statement of the zeroth law follows the existence of an intensive
variable, the temperature, which serves to predict the behaviour of
two initially isolated systems at equilibrium when they are brought
into thermal contact. It states that the equality of the temperature
parameter is a necessary and sufficient condition that no irreversible
heat exchange will occur.

It is of considerable interest to fully understand the possible
limitations of such a basic law. In the following, we shall show that
systems with negative specific heat can indeed violate the zeroth law.
This is more than just an academic concern: systems with negative
specific heat do exist in Nature. Although it is readily shown that
systems well described by the canonical ensemble, such as those which
are weakly coupled to an environment at fixed temperature, cannot have
negative specific heat, others may. In particular, if a system is
entirely isolated, that is, if it is in the microcanonical ensemble,
the specific heat may well become negative \cite{gross}.  To this end,
however, it is obviously necessary that the microcanonical ensemble
lead to significantly different predictions from the canonical
one. This cannot happen for large systems with short-range
interactions.  However, if either the interactions are long-range or,
perhaps more realistically, if the system is {\em small} with respect
to the interaction range, then it is not possible to refer to general
statements concerning the equivalence of ensembles and negative
specific heats can be, and indeed have been, observed. Such systems
include among others gravitational systems (long-range interactions)
and atomic clusters (small systems), see \cite{lyndenbell68,
thirring, aronson72, lyndenbell77, lydenbell, laliena, chavanis,
barre01, cohen, labastie, nava, schmidt, poschA}.  Analytically,
mean-field models, involving
interactions of all particles with all, are well-known to display
negative specific heats \cite{barre01, campa06} in the microcanonical
ensemble for parameters which, in the canonical ensemble, would
correspond to the vicinity of a tricritical point \cite{francois}.

Since systems with negative specific heat are thermodynamically
unstable when they are thermally coupled to the surrounding medium
\cite{reichl}, anomalous behaviour is surely to be expected when such
systems interact. However, it is not apparent that this will cause
violations of the zeroth law. The reason is precisely that the zeroth
law always involves coupling between systems. Heat exchange is thus
always allowed whenever the zeroth law is tested, and one might hence
think that this always leads to a canonical-like case, in which the
restriction on fixed energy is lifted and negative specific heats
cannot occur. We shall show that this is not the case.

In order to test the zeroth law for small systems in the
microcanonical ensemble, one also faces yet another issue: for such
systems it is often not straightforward how to define the temperature
posited by the zeroth law. Under these circumstances, how, then, shall
we know whether two systems really have the same temperature or not?

A simple way around the last question is straightforward: it suffices
to take two identical systems, that is, two systems having the same
Hamiltonian with the same energy, volume and number of particles, and
then to couple them weakly. Under these conditions, whatever the
temperature is in each system, it certainly will be the same for both;
yet irreversible changes may occur in the composite system even under
such circumstances.

To be specific, let us focus on the following Hamiltonian describing a
system of $N$ classical XY-rotors with phases $\theta_{i}$, defined by
\begin{eqnarray}
 H&=&\sum_{i=1}^{N} \dfrac{p_{i}^{2}}{2} + \dfrac{J}{2
N}\sum_{i,j=1}^{N} \left(1-\cos\left( \theta_{i}-\theta_{j}\right)
\right) \nonumber\\ &&-K \sum_{i=1}^{N} \cos(\theta_{i+1}-\theta_{i})
, \label{hamiltonian}
\end{eqnarray}
where $J>0$ is the global ferromagnetic coupling and $K$ is the
nearest neighbors coupling; it can be negative or positive. The rotors
are placed on a one-dimensional lattice with periodic boundary
conditions.  When $K=0$, Eq.~(\ref{hamiltonian}) is the classic
Hamiltonian Mean Field model introduced in~\cite{antoni}, which has a
second order phase transition at the critical temperature
$T_{c}=J/2$~\cite{antoni} in the canonical ensemble (though the
microcanonical and the canonical ensembles are equivalent when
$K=0$). This phase transition is characterized by the behavior of the
magnetization order parameter $m$,
\begin{equation}
m=\frac{1}{N} \sqrt{\left( \sum_{i=1}^{N}\cos
\theta_{i}\right)^{2}+\left( \sum_{i=1}^{N}\sin
\theta_{i}\right)^{2}}. \label{order}
\end{equation}
For $K<0$ it was shown~\cite{campa06} that this system shows
inequivalence of ensembles for an interval of values of
$K$. Specifically, there exists a region in which this system shows
negative specific heat in the microcanonical ensemble.

\begin{figure}[t]
\begin{center}
\scalebox{0.4}{\includegraphics*[]{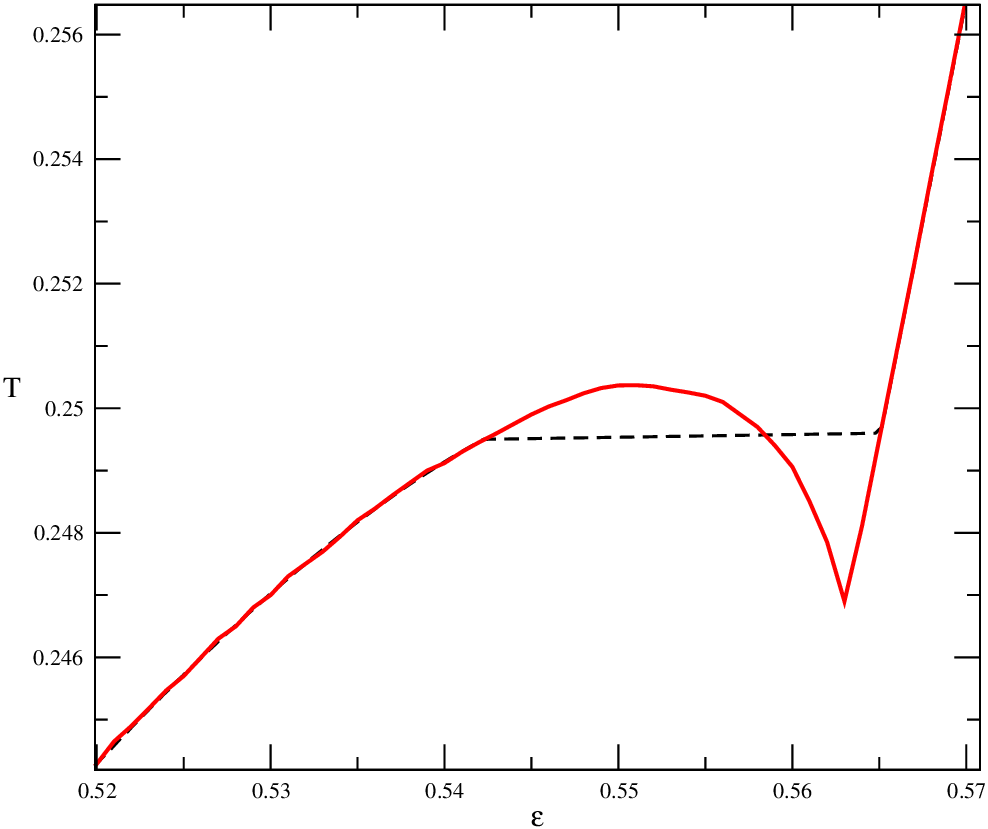}}
\end{center}
\caption{\label{caloric}\footnotesize{} Caloric curves ($T$ vs
$\varepsilon$) in the canonical (black) and microcanonical (red)
ensemble for the model (\ref{hamiltonian}) obtained from the exact
expressions (\ref{freecanonical}) and (\ref{entropymicro}),
$K=-0.178$.}
\end{figure}

The canonical equilibrium properties are determined by the free energy
per particle $f(\beta)$~\cite{campa06} ($J=1$),
\begin{eqnarray}
 -\beta f(\beta)&=&\max_{m} \left[ \frac{1}{2}\ln \dfrac{2\pi}{\beta}
+ \ln \lambda(\beta m,\beta K) \right. \nonumber\\ &&-
\left. \dfrac{\beta(1+m^{2})}{2}\right], \label{freecanonical}
\end{eqnarray}
where $\lambda(z,\alpha)$ is the largest eigenvalue of the transfer
matrix given by the following operator
\begin{eqnarray}
 (\widehat{T}\varphi)(\theta)&=&\int d\theta^{'} \exp\left[
\dfrac{1}{2}z(\cos\theta+\cos\theta^{'})\right.  \nonumber\\ &&+
\left. \alpha \cos(\theta-\theta^{'})\right]
\varphi(\theta^{'}). \label{operator}
\end{eqnarray}
In the microcanonical ensemble the thermo\-dy\-namic pro\-perties are
given by the entropy per particle $s_{\mu}$ that is obtained from
(\ref{freecanonical}) by using the mean-field formalism introduced in
Ref.~\cite{francois},
\begin{eqnarray}
 s_{\mu}(\varepsilon)&=&\max_{m} \min_{\beta}\left[\beta \varepsilon +
\frac{1}{2}\ln\dfrac{2\pi}{\beta} + \ln \lambda(\beta m,\beta K)
\right. \nonumber\\ &&- \left. \dfrac{\beta(1+m^{2})}{2} \right],
\label{entropymicro}
\end{eqnarray}
where $\varepsilon$ is the energy per particle. The canonical entropy
$s_{c}$ is found from (\ref{freecanonical}) by a Legendre
transformation with respect to $\beta$ and has therefore the usual
concavity property. The same does not hold, however, for
$s_{\mu}(\varepsilon)$ as is seen in figure~\ref{caloric}.

When we now couple both systems, the full system is still isolated
and the total energy $E=E_{1}+E_{2}+E_{int}$ is constant.  As
we consider weak couplings, the interaction energy can be neglected,
so the total energy is the sum of the individual energies of each
subsystem. Under this condition the total entropy per particle can be
written as
\begin{equation}
 s(\varepsilon_{1},\varepsilon_{2})=\frac{1}{2} \left[
 s_{\mu}(\varepsilon_{1})+s_{\mu}(\varepsilon_{2})\right] ,
\end{equation}
with $s_{\mu}$ given by (\ref{entropymicro}) and the restriction
$\varepsilon_{2}=2\varepsilon-\varepsilon_{1}$. Here
$\varepsilon_{\gamma}$ is the energy density of subsystem $\gamma$ and
$\varepsilon$ is the energy density of the full system, which is
constant. The correct description of the full system is thus given by
the microcanonical entropy, although the energy conservation
constraint was removed for each individual subsystems. By the second
law
of thermodynamics, the equilibrium state
$(\varepsilon_{1}^{*},\varepsilon_{2}^{*}, m_{1}^{*},m_{2}^{*})$ will
be such that the total entropy will be maximum. So, the total entropy
per particle is given by
\begin{eqnarray}
s(\varepsilon)=\max_{\varepsilon_{1}}\frac{1}{2}\left[s_{\mu}(\varepsilon_{1})+s_{\mu}(2\varepsilon-\varepsilon_{1})
\right]=\max_{\varepsilon_{1}} s(\varepsilon,\varepsilon_{1}).
\label{entropycoupled}
\end{eqnarray}
From this optimization problem we obtain the following condition for
the maximun entropy:
$T_{\mu}^{1}(\varepsilon_{1}^{*})=T_{\mu}^{2}
(2\varepsilon-\varepsilon_{1}^{*})$,
where $\varepsilon_{1}^{*}$ is the energy that maximizes the total
entropy and $T_{\mu}^{\gamma}(\varepsilon_{\gamma})$ is the
temperature of the subsystem $\gamma$ defined as the 
inverse derivative
of the entropy with respect to the energy at constant $N,V$.

\begin{figure}[t]
\begin{center}
\scalebox{0.4}{\includegraphics*[]{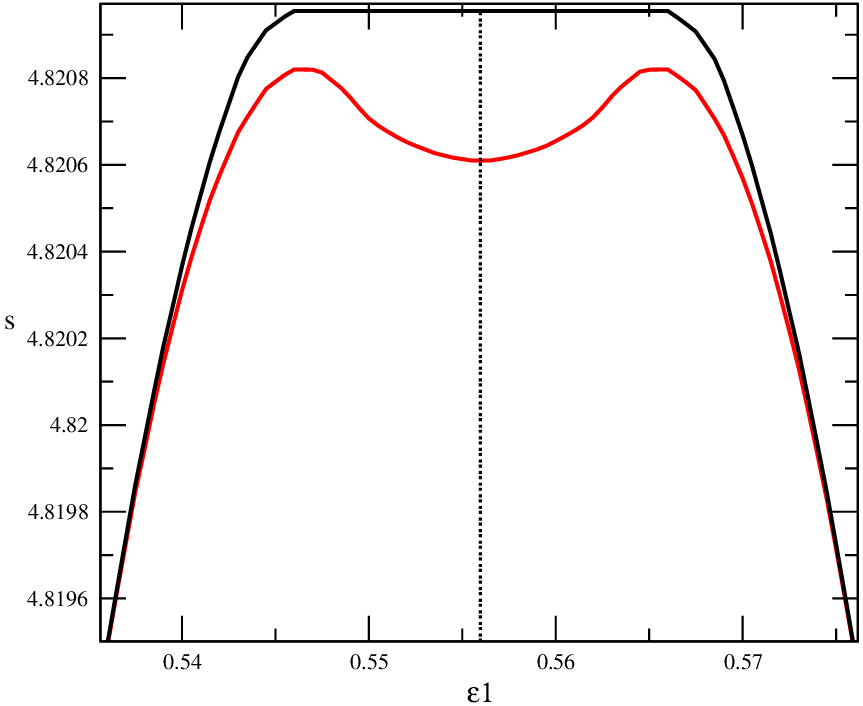}}
\end{center}
\caption{\label{entropy}\footnotesize{} Total entropy per particle,
$s(\varepsilon,\varepsilon_{1})$, for the coupled subsystems canonical
(black) and microcanonical (red), the dotted line indicates the value
of energy density before coupling, for identical subsystems with
negative specific heat. The values of the parameters are
$K_{1}=K_{2}=-0.178,\varepsilon=0.55597$.}
\end{figure}

By using (\ref{entropymicro}) and (\ref{entropycoupled}) we can
predict the equilibrium values of magnetization, energy density and
temperature of each subsystem.  In figure~\ref{entropy} we show
$s(\varepsilon,\varepsilon_{1})$ vs $\varepsilon_{1}$, microcanonical
and canonical. For the microcanonical case there are two
maxima. Because both subsystems are identical, in principle which
subsystem evolves to the higher energy is random. The total entropy
before the coupling is now a minimum, that is, an unstable
state.  That this can occur was already noted in \cite{lyndenbell77}.
Therefore, the state of maximum entropy is now obtained when the
subsystems have different values of magnetization and energy density,
i.e. when two \textit{phases} appear (a similar phenomenon was
observed in \cite{poschB}). The total entropy is therefore increased
irreversibly if two subsystems with the same intensive parameters with
negative specific heat are thermally coupled. Note that the final
temperatures of both subsystems are identical, but in general differ
from the initial temperature \cite{future}. This difference is often
too small to be detected numerically but its existence is rigorously
established by the analytical results.

To confirm our results we have performed numerical simulations of the
complete system using a fourth-order symplectic algorithm with a time
step 0.1~\cite{atela}. We run the simulations for a time interval
$\tau_{eq}$ to let both subsystes reach equilibrium without
interaction (i.e. $\eta=0$). Once they have equilibrated, we increase
the coupling linearly during a time interval $\tau_{a}$, after which
the coupling is maintained constant, $\eta>0$.

We choose parameters in the region in which the uncoupled subsystems
exhibit negative specific heat, and both subsystems have the same
values of all their parameters initially: energy per particle
($\varepsilon_{1}=\varepsilon_{2}$), magnetizations ($m_{1}=m_2$) as
well as particle numbers ($N_{1}=N_{2}$). From these equalities it
follows that the derived variables, temperatures ($T$) and specific
heats ($c$) are identical as well.  The value of these thermodynamic
variables are defined through the microcanonical entropy:
$T(\varepsilon)=(ds_{\mu}(\varepsilon)/d\varepsilon)^{-1}$ and
$c(\varepsilon)=-(d^{2}s_{\mu}(\varepsilon)/d\varepsilon^{2})^{-1}
(ds_{\mu}(\varepsilon)/d\varepsilon)^{2}$. The value of the
magnetization is such that the entropy for $\varepsilon$ fixed is
maximum, therefore $m=m(\varepsilon)$ as well.  Note further that for
Hamiltonians with a quadratic kinetic energy, as (\ref{hamiltonian}),
we may without appreciable error evaluate the temperature using the
expression for the kinetic temperature, which is given in this case as
twice the mean kinetic energy per particle, since the finite $N$
microcanonical corrections are negligible. This identification was
verified by comparing with the momentum distribution, which we found
to be Maxwellian. The measurement of the temperature in the simulation
is therefore straightforward in this case.

Here we report the effect of the following choice for the 
coupling Hamiltonian
\begin{equation}
 H_{int}^{p}=\eta \sum_{i=1}^{N_{int}} p_{i}^{1}p_{i}^{2} ,
\label{interP}
\end{equation}
where $p_{i}^{\gamma}$ is the momentum of the $i-$th particle of
subsystem $\gamma$. The advantage of this apparently peculiar choice
is that this coupling does not exert a direct influence on the
magnetizations of the two subsystems.

\begin{figure}[t]
\begin{center}
\scalebox{0.4}{\includegraphics*[]{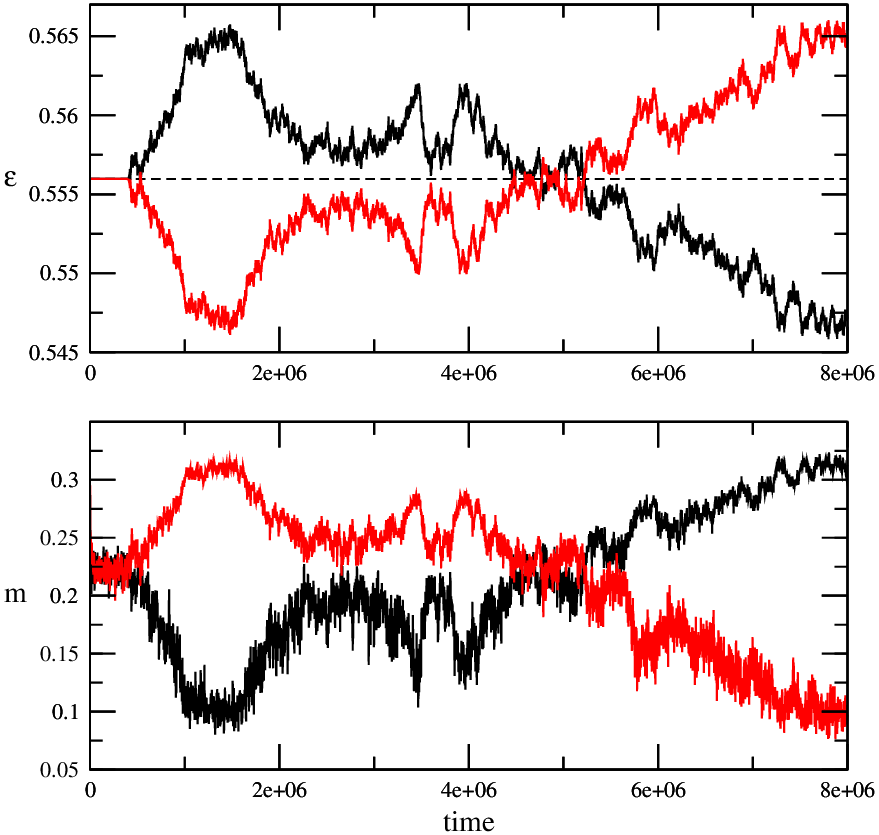}}
\end{center}
\caption{\label{energy_Pc}\footnotesize{} Temporal evolution of energy
density (top) and magnetization (bottom) of subsystem 1 (black) and
subsystem 2 (red), using the coupling $H_{int}^{p}$. The curves
were obtained by averaging over a sliding time window. 
The values of parameters are $K_{1}=K_{2}=-0.178,
N_{1}=N_{2}=5000, \eta=0.1,
N_{int}=10,\varepsilon_{1}^{0}=\varepsilon_{2}^{0}=0.55597$.}
\end{figure}

In Figure \ref{energy_Pc} we show the behaviour of the internal energy
and the magnetization of both subsystems. Clearly, they do not remain
at their initial equilibrium values; rather, due to finite size
effects, the systems jump between the two degenerate equilibrium
states of the coupled system \cite{Katz}: thus either subsystem one is
in a magnetized state and subsystem two in an $m=0$ state, or
viceversa. The oscillation does indeed roughly take place between the
two equilibrium values of the magnetization, which for the parameter
values used here can be computed, along the lines described above, to
be $T_{\mu}^{1}=T_{\mu}^{2}=0.25, \varepsilon_{1}^{*}=0.56594,
\varepsilon_{2}^{*}=0.546, m_{1}^{*}=0, m_{2}^{*}=0.32$. There is a
problem, however, with the fact that neither subsystem ever comes
reasonably close to being truly paramagnetic ($m=0$)\cite{n_finite}. As it turns out,
this is also a finite-size effect.  As is readily seen, the mean-field Ising model {\em at\/} criticality
satisfies
\begin{equation}
    \langle m^2\rangle=const.\cdot N^{-1/2}
    \label{eq:1}
\end{equation}
which states that the typical magnetization fluctuations at, or in the
vicinity of, a second order phase transition are of order
$N^{-1/4}$. A study of the $N$-dependence of the magnetizations
observed in our system is compatible with this scaling behavior
\cite{future}.  Thus, these large fluctuations due to the vicinity
of a second order phase transition, which is known to exist at
$\varepsilon_{c}\approx0.5633$, are the reason why the zero
magnetization phase is not clearly observed in our finite systems.

The reason for the appearance of two phases lies, of course, in the
thermodynamic instability of systems with negative specific heat, as
discussed previously. Such systems can only exist through the
existence of constraints, such as the conservation of energy, which
keeps them from relaxing to a more probable state. What we have
therefore shown is simply that coupling two such systems with each
other provides enough freedom for an irreversible relaxation process
to take place.  It can also be shown \cite{future} that this
phenomenon is not limited to the specific case studied here: any
system with negative specific heat will show similar behaviour. The
type of coupling used is also unimportant.

Further, it is possible to use a third system (with positive specific
heat) as a thermometer to measure the temperature of the systems with
negative specific heat. Thus, we can check that our systems are in a
stable equilibrium with the thermometer at the same temperature before
we put them in thermal contact. There exists, however, a condition
over the third system: it needs to have a \textit{small heat
capacity}, otherwise it may allow large enough energy fluctuations to
drive the system with negative specific heat out of the microcanonical
ensemble (this condition on the thermometer was already
discussed in \cite{thirring,lyndenbell77,lydenbell}; explicit
simulations of this situation will be presented in \cite{future},
where we use an ideal gas as the thermometer).

Summarizing, we have shown how coupling two systems with negative
specific heat leads to an irreversible change in the intensive
variables of the subsystems. In the simulations, the entire
system does not settle down to a well- defined state. Rather, it
displays slow oscillations between two degenerate states of
thermodynamic equilibrium due to the finite size of the
systems. It should further be noted that neither of these states
corresponds to the canonical equilibrium state.

Since both subsystems began at equilibrium with the same intensive
parameters, the validity of the zeroth law implies that the coupling
should not produce any noticeable effect. Since, as we have seen both
through an analytical approach and by explicit numerical work, the
system in fact relaxes to an inhomogeneous state in which the two
weakly coupled subsystems have widely different values of the
magnetization and of the internal energy, we have unambiguously
observed a instance of a violation of the zeroth law of
thermodynamics.





\begin{thebibliography}{99}
\bibitem{n_finite} Negative specific heat may arise in systems that are
of \textit{mean-field type} as defined in~\cite{francois}, near the
tricritical point. For this kind of systems we can write a Landau-like
free energy density and we can approximate it by
$\phi=a_{2}m^{2}+a_{4}m^{4}+a_{6}m^{6}$. If the system is near the
tricritical point $(a_{2}\approx0,a_{4}\approx0)$ then $m\sim
N^{-\alpha}$ with $\alpha=1/6$, but if the system is near a critical
point, $a_{2}\approx0$ and $a_{4}$ is non zero, then $\alpha=1/4$. In
either case, for very large $N$ the normal $\alpha=1/2$ behaviour is
recovered.
\end{thebibliography}
\end{document}